# Phonon-driven wavefunction localization promotes room-temperature, pure single-photon emission in large organic-inorganic lead-halide quantum dots

Leon G. Feld,[1,2,3] Simon C. Boehme,[1,2] Sebastian Sabisch,[1,2] Nadav Frenkel,[4] Nuri Yazdani,[5] Viktoriia Morad,[1,2] Chenglian Zhu,[1,2] Mariia Svyrydenko,[1,2] Rui Tao,[1,2] Maryna Bodnarchuk,[1,2] Gur Lubin,[4] Miri Kazes,[4] Vanessa Wood,[5] Dan Oron,[4,*] Gabriele Rainò[1,2,3,*] and Maksym V. Kovalenko[1,2,3,*]

[1] Institute of Inorganic Chemistry, Department of Chemistry and Applied Biosciences, ETH Zürich, CH-8093 Zürich, Switzerland

[2] Laboratory for Thin Films and Photovoltaics, Empa – Swiss Federal Laboratories for Materials Science and Technology, CH-8600 Dübendorf, Switzerland

[3] National Centre of Competence in Research (NCCR) Catalysis, ETH Zürich, CH-8093 Zürich, Switzerland

[4] Department of Molecular Chemistry and Materials Science, Weizmann Institute of Science, Rehovot 76100, Israel

[5] Department of Information Technology and Electrical Engineering, ETH Zürich, CH-8093 Zürich, Switzerland

Emails: mvkovalenko@ethz.ch, rainog@ethz.ch, dan.oron@weizmann.ac.il

**ABSTRACT**

In lead halide perovskites (APbX$_3$), the effect of the A-site cation on optical and electronic properties has initially been thought to be marginal. Yet, evidence of beneficial effects on solar cell performance and light emission is accumulating. Here, we report that the A-cation in soft APbBr$_3$ colloidal quantum dots (QDs) controls the phonon-induced localization of the exciton wavefunction. Insights from ab initio molecular dynamics and single-particle fluorescence spectroscopy demonstrate that anharmonic lattice vibrations and the resulting polymorphism act as an additional confinement potential. Avoiding the trade-off between single-photon purity and optical stability faced by downsizing conventional QDs into the strong confinement regime, dynamical phonon-induced confinement in large organic-inorganic perovskite QDs enables bright (10$^6$ photons/s), stable (> 1h), and pure (> 95%) single-photon emission in a widely tuneable spectral range (495-745 nm). Strong electron-phonon interaction in soft perovskite QDs provides an unconventional route toward the development of scalable room-temperature quantum light sources.

**MAIN**

Reliable and scalable single-photon sources (SPSs) are essential for the broad adoption of quantum optical methods in computing,[1] communication[2] and imaging.[3] So far, epitaxially-grown quantum dots (QDs) operated at liquid-helium temperatures excel with on-demand photon generation at exceptional single-photon purity, fluorescence rates, coherence times and indistinguishability.[4] Recently, coherent,[5] indistinguishable[6] and fast single-photon emission[7] was also demonstrated for colloidal cesium lead halide perovskite QDs at similar temperatures. Yet, capitalizing on the excellent performance of such sources is constrained by the need for cryogenic operation that limit their scalability and accessibility. Room-temperature SPSs alleviate the need for cryogenic cooling thus reducing significantly the system complexity. Amongst other room-temperature SPSs like doped organic crystals[8] or defect-based emitters,[4] colloidal QDs exhibit scalable and low-cost synthesis based on simple solution chemistry, precise control over their optical properties, and ease of solution processing. Comprehensive structural and compositional engineering elevated the photoluminescence (PL) characteristics of colloidal QDs to near-unity quantum yields (QY), suppressed PL fluctuations and narrow-band emission.[9-11] Already implemented in various classical optoelectronic technologies,[12-14] some of which are widely commercialized, colloidal QDs are becoming promising candidates for single-emitter applications.[15-17] However, colloidal QDs are yet to match the performance of cryogenically-cooled epitaxial QDs in terms of single-photon purity and PL stability over time.

For colloidal QDs, improvements to their single-photon purity can be obtained by reducing their physical size.[18-20] Such size-confinement yields discretized energy levels, strong (multi-)exciton interaction and accelerated non-radiative Auger-Meitner[21] recombination of multiexcitons.[22,23] The resulting single-photon purity (exceeding 90% in colloidal QDs)[17,18,24] is inferred from the amplitude of the dip in $g^2(\tau)$, the second-order photon-photon correlation function, at zero-delay time (Figure 1a). However, enhancing the single-photon purity via QD downsizing limits the spectral tunability and comes at a high cost: Firstly, vulnerability to surface defects and matrix effects that cause stochastic intensity fluctuations ("blinking") and photobleaching is amplified in small QDs.[9,10,25] Secondly, increased coupling to surface vibrations broadens the emission spectrum of small QDs.[9,26] Finally, the volume-scaling of the absorption cross sections translates into the need for high excitation densities, eliminating the potential for on-chip optical pumping by LEDs and prompting uncorrelated background photons that contaminate single-photon emission. In conclusion, finding alternative ways toward high single-photon purities without the need for detrimental size confinement would improve the quality, stability, and versatility of colloidal QDs as room-temperature SPSs.

In search for alternatives to size confinement, we identified the coupling between electronic and vibrational degrees of freedom (vibronic coupling)[27] as an underutilized route towards confining charge carriers in semiconductor sources. Vibronic coupling broadens the emission lines of semiconductors employed in display and lighting applications, limits the overall charge mobility in crystalline and amorphous materials used in photovoltaics, and causes the loss of exciton optical coherence, a key resource for photon-based quantum computing. Nevertheless, while generally considered a detrimental interaction, vibronic coupling can mediate fascinating phenomena, such as Bose-Einstein condensation of exciton-polaritons,[28] super-conductivity,[29] and optical cooling.[30]

In crystalline materials at finite temperatures, structural disorder is incited by anharmonic lattice vibrations (anharmonic phonons) that displace atoms from their average lattice site (Figure 1b) and is translated into electronic disorder through vibronic coupling.[31] Other sources of disorder include lattice site disorder in defected crystalline solids[32] or structural disorder in amorphous solids.[33] The random potential introduced by the disorder can serve as an effective confinement potential that localizes electronic wavefunctions.[34-36] Therefore, we hypothesize that the combination of phonon-induced disorder and exciton-phonon coupling can act in addition to the quantum size effect and thereby enhance quantum confinement of (multi-)excitons and consequently single-photon purity (Figure 1c). This approach does not come with the risk of trap state formation which usually reduces PL quantum yield (PLQY) and the overall performance of single-photon sources, but necessarily requires the use of materials exhibiting high amplitude ('anharmonic') vibrations at room temperature.

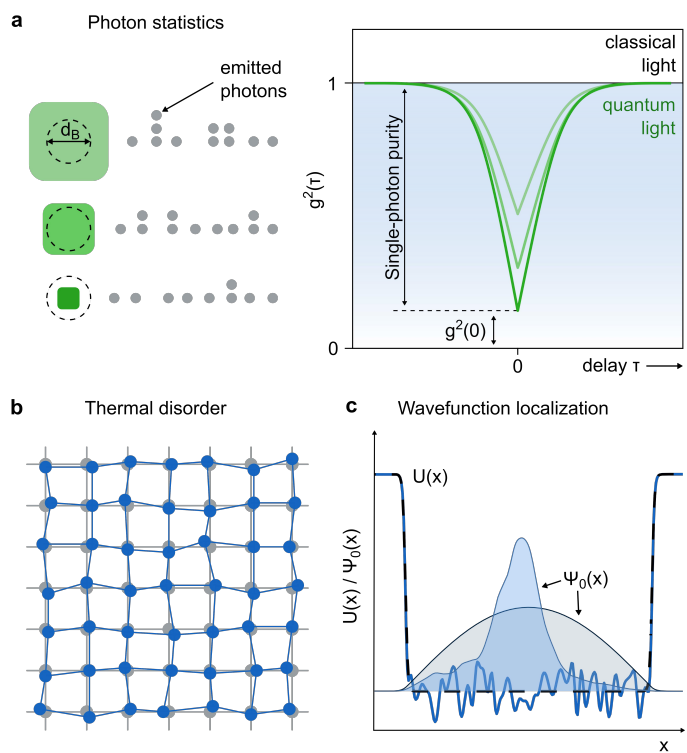

**Figure 1.** Single-photon emission through conventional and disorder-induced quantum-confinement. (a) Illustration of the emitted photon statistics from individual colloidal quantum dots and its size-dependence. Grey dots indicate the photons emitted from particles with different particle sizes relative to their Bohr diameter ($d_B$). Photon statistics are evaluated via the second-order autocorrelation of the photon arrival times ($g^2(\tau)$). Single-photon purity is inversely proportional to $g^2(\tau)$ at a time delay ($\tau$) of zero. (b) Schematic drawing of disorder through random atomic displacement (blue) from the equilibrium positions (grey). (c) Illustration of weak disorder-induced wavefunction localization in a QD. The potential energy surface $U(x)$ of a QD without disorder (black dashed line) results in a probability density distribution $\Psi_0(x)$ delocalized across the entire QD (grey area). A disordered potential (blue line) induces localization of the probability density distribution (blue area).

A suitable platform to test our hypothesis is provided by lead halide perovskite (LHP) materials ($APbX_3$; A=FA, MA, Cs; X=Cl, Br, I) that emerged as promising solar cell materials due in part to their high defect-tolerance and facile solution processing. Colloidal perovskite QDs exhibit high defect-tolerance and therefore achieve near-unity PLQY and relatively weak blinking without the need for delicate core-shell engineering.[37,38] Perovskite QDs have demonstrated single-photon emission at room-temperature[39] as well as bright, coherent, and indistinguishable single-photon emission at cryogenic temperatures.[5,6] Initially, the A-site cation has largely been deemed a bystander in defining the photo-physics of LHPs due to its lack of direct electronic contribution to the band edge states.[40] However, A-site cations can influence the structure of the $PbX_3^-$ sublattice and associated lattice dynamics, which, via a strong coupling of the exciton to anharmonic lattice vibrations, also impact the optoelectronic performance.[41,42] Consequently, the A-site cation's importance for optical and electronic characteristics in LHPs is being recognized and exploited to improve, for example, LHP solar cell performance.[40,43-48]

In this work, we suggest that the large dynamic structural disorder in $FAPbBr_3$ QDs increases the effective quantum confinement and hereby renders even large QDs high-quality quantum emitters of single photons. Identifying disorder as an alternative route towards quantum confinement alleviates well-known size-dependent QD performance trade-offs and extends single-photon emission capability to QDs of sizes larger than the Bohr diameter, hereby straightforwardly improving both the source brightness and its stability. Ab-initio molecular dynamics (AIMD) simulations pinpoint A-site-cation-controlled anharmonic vibrational modes and associated polymorphism as the origin of the structural disorder responsible for dynamic wavefunction localization and enhanced quantum confinement. The enhanced Auger-Meitner recombination of multi-excitons even in large $FAPbX_3$ QDs enables pure single-photon emission, which is also bright, stable, and finely spectrally tuneable across the visible spectrum.

We selected $CsPbBr_3$ and $FAPbBr_3$ QDs as candidates to study the effect of the A-site cation on structural disorder because their room-temperature crystal structures and dynamics are tuned by the choice of the A-cation. The room-temperature crystal structure of $CsPbBr_3$ is orthorhombic as determined by single-crystal X-ray diffraction (XRD; Figure 2a). On the other hand, $FAPbBr_3$ adopts an average cubic crystal structure (Figure 2b). Figure 2c and d display magnified views of the Pb-Br-Pb bonding of the corner-shared PbBr$_6$ octahedra. CsPbBr$_3$ and FAPbBr$_3$ differ in the Pb-Br-Pb angles as well as thermal displacement ellipsoids that describe the static and dynamic disorder occurring at finite temperatures. In CsPbBr$_3$, the average Pb-Br-Pb angle deviates from 180° as octahedral tilting compensates for the undersized Cs-cation. Contrarily, a larger cation size and thus better size fitness of FA in the octahedral void stabilizes an average cubic structure, as evidenced by a Pb-Br-Pb angle of 180°. The shape of the thermal displacement ellipsoids of Br reveals that in both systems octahedral tilting dominates disorder in the PbBr$_3^-$ sublattice, albeit the cubic FAPbBr$_3$ structure additionally exhibits stronger displacement along the Pb-Br bond.

Single-particle PL spectroscopy can optically probe the small changes in structure and dynamics of the PbBr$_3^-$ sublattice that are introduced by the different A-site cations. Figure 2e displays representative PL spectra of CsPbBr$_3$ QDs and FAPbBr$_3$ QDs with edge lengths of 9.9(1.2) nm as determined by transmission electron microscopy (Figure S1). With relative edge lengths of 1.61(19) and 1.27(16) times their Bohr diameter (d$_B$; 6.16 nm CsPbBr$_3$ and 7.76 nm for FAPbBr$_3$[49]), the samples should display similar size-induced quantum confinement. Samples were prepared following recently published room-temperature procedures (details in the Supporting Information).[37] Compared to CsPbBr$_3$, the spectrum of FAPbBr$_3$ exhibits a peak center that is redshifted by 72 meV and an increase in PL peak width (FWHM) by 13 meV. The red-shift is associated with the straightened and stretched Pb-Br-Pb bonding that reduces the orbital overlap (Figure 2a-d), and the increased peak width evidences the stronger electron-phonon coupling in FAPbBr$_3$.[26,45] Moreover, we observed an exponential low energy (Urbach) tail in single-particle PL spectra, which are displayed in Figure 2f. For comparable QD sizes and across a large size range, the tails are consistently steeper for CsPbBr$_3$ than in FAPbBr$_3$ QDs (Figure 2g), suggesting, as elucidated in a prior study,[45] stronger exciton-phonon coupling. Stronger exciton-phonon coupling in FAPbBr$_3$ QDs is moreover confirmed by several works probing the exciton phonon replica at cryogenic temperatures.[50,51] Urbach tails are associated with exciton-phonon coupling through the formation of localized states that result from thermal disorder and form the low-energy tail.[33,35] They thus also serve as a first indication of the hypothesized wavefunction confinement which may be enhanced in pseudo-cubic FAPbBr$_3$ QDs.

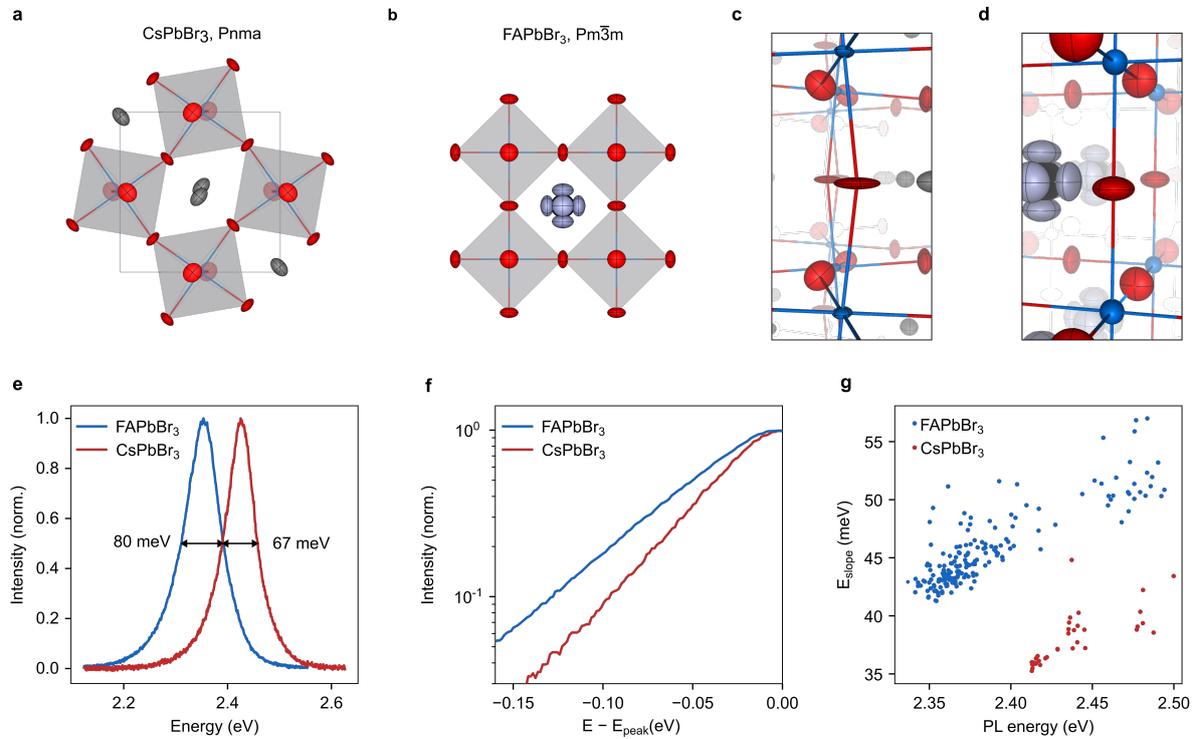

**Figure 2.** Effect of A-cation on spectral features of perovskite QDs. (a,b) Room-temperature crystal structures of CsPbBr$_3$ (a) and FAPbBr$_3$ (b) obtained by single-crystal X-ray diffraction (red: Br, blue: Pb, grey: Cs, purple: N, black: C). (c,d) Magnified view of the Pb-Br-Pb bond and thermal displacement ellipsoids in the crystal structures of CsPbBr$_3$ (c) and FAPbBr$_3$ (d). (e) Photoluminescence spectra of single FAPbBr$_3$ and CsPbBr$_3$ QDs with similar size (both 9.9(1.2) nm) displaying a spectral redshift and spectral broadening upon replacing Cs by FA. (f) Red tail of the PL spectra from single FAPbBr$_3$ and CsPbBr$_3$ QDs shifted by their PL center. (g) Slope energy as a function of PL center obtained by fitting eq (S1) to the red tail PL spectra of single FAPbBr$_3$ and CsPbBr$_3$ QDs of various sizes.

An understanding of electron-phonon coupling can be gleaned from finite-temperature ab-initio molecular dynamics (AIMD) simulations based on density functional theory (DFT). We performed simulations for 3.6 nm large ABr-terminated $CsPbBr_3$ and $FAPbBr_3$ QD models at the Perdew-Burke-Ernzerhoff level of theory.[52] Previously, such a balance between computationally approaching experimental QD sizes yet ensuring a sufficiently high level of theory has successfully reproduced experimental results relating to surface chemistry[53] and exciton-phonon interaction.[7,26,54,55] To study wavefunction confinement, we computed the wavefunction of the highest occupied molecular orbital (HOMO) which represents the hole wavefunction. Figure 3a displays representative snapshots of the HOMO wavefunctions at 10 K and 300 K. At low temperatures, the wavefunctions fully delocalize across the entire QD volume for both $CsPbBr_3$ and $FAPbBr_3$ QDs. At 300 K, however, the wavefunction strongly contracts and localizes in the $FAPbBr_3$ QD, while such a thermally activated wavefunction confinement is considerably weaker in the $CsPbBr_3$ QD.

Figure 3b shows the average wavefunction sizes in the $CsPbBr_3$ and $FAPbBr_3$ QDs along AIMD trajectories at 10, 100, and 300 K. In both materials, the wavefunction sizes remain between 1.6 and 1.8 nm at 10 and 100 K, limited by QD size. Above 100 K the wavefunction size of the $FAPbBr_3$ QD decreases to just 0.93 ± 0.17 nm, whilst that of the $CsPbBr_3$ QD only decreases to 1.40 ± 0.05 nm. This observation is consistent across several QD diameters (Figure S2) and was recently even observed in bulk $CsPbBr_3$.[55,56] Moreover, the wavefunction size for $FAPbBr_3$ QDs at 300 K is nearly independent of QD diameter, implying a disorder-limited wavefunction extension (Figure S2).

Importantly, the wavefunction localization is dynamic. Both the size and location inside the QD vary between snapshots (Supporting video, Figure S3). Moreover, the time-averaged wavefunction, integrated across the entire AIMD trajectories (Figure S4), smoothly covers the entire QD volume which excludes static contributions to the localization. The highly dynamic wavefunction as well as the temperature dependence suggest that anharmonic phonons are the primary cause for the transient crystal disorder and the associated wavefunction localization. This mechanism is reinforced by temperature-dependent radial and angular distribution functions that support the existence of temperature-induced atomic displacement from their average position (Figure S5,6). Moreover, the structural disorder caused by these atomic displacements is discernible from temperature-dependent spatial correlation functions of the $PbBr_3^-$ sublattice (Figure S7). Finally, the correlation of temperature-induced structural disorder and wavefunction localization with the population of phonon modes underpins allocating the disorder to phonons (Figure S8).

Further insight into the phonon-driven wavefunction localization is provided by the autocorrelation function of the HOMO wavefunction coefficients which describes the time-evolution of the wavefunction. Figure 3c depicts the normalized wavefunction autocorrelation functions at 300 K exhibiting decays within hundreds of femtoseconds that envelop oscillations with a period of roughly 200 fs. Notably, the loss of correlation is significantly faster for the $FAPbBr_3$ QD than for the $CsPbBr_3$ QD.[7,55] This observation resembles the bandgap energy autocorrelations, which have previously been assessed to derive the phonon-driven electronic dephasing in these and other semiconductor materials.[26,54]

The power spectra of the wavefunction autocorrelation in the inset in Figure 3c reveal the dominant vibrational features driving the wavefunction localization in both systems. While the wavefunction-phonon coupling strength in $CsPbBr_3$ essentially concentrates within a single peak at 17 meV, the coupling strength in $FAPbBr_3$ QD additionally derives from a broad range of low-energy phonons. The population of these modes is confirmed by vibrational spectral density at these energy in our AIMD simulations as well as experimental Raman spectra with good agreement between the two methods (Figure 3d). Whilst the 17 meV peak is associated with a stretching mode along the Pb-Br bonds, the highly anharmonic tilting of $PbBr_6$ octahedra contributes to the lower-energy modes.[50,57] Peculiar to average cubic perovskites like $FAPbBr_3$, this anharmonic motion induces local symmetry breaking evoking polymorphic crystals prompting the labelling as "pseudo-cubic".[58,59] Such polymorphism leads to significant bandgap renormalization and is hence also expected to strongly affect the wavefunction localization of the excitons.[59] Consequently, the ability to tune between the orthorhombic and pseudo-cubic structure through the A-cation allows us to manipulate the polymorphism and the resulting wavefunction localization. Based on structural similarities,[60] we expect other pseudo-cubic perovskites like MAPbBr$_3$ or AzPbBr$_3$ to exhibit similarly pronounced wavefunction localization.

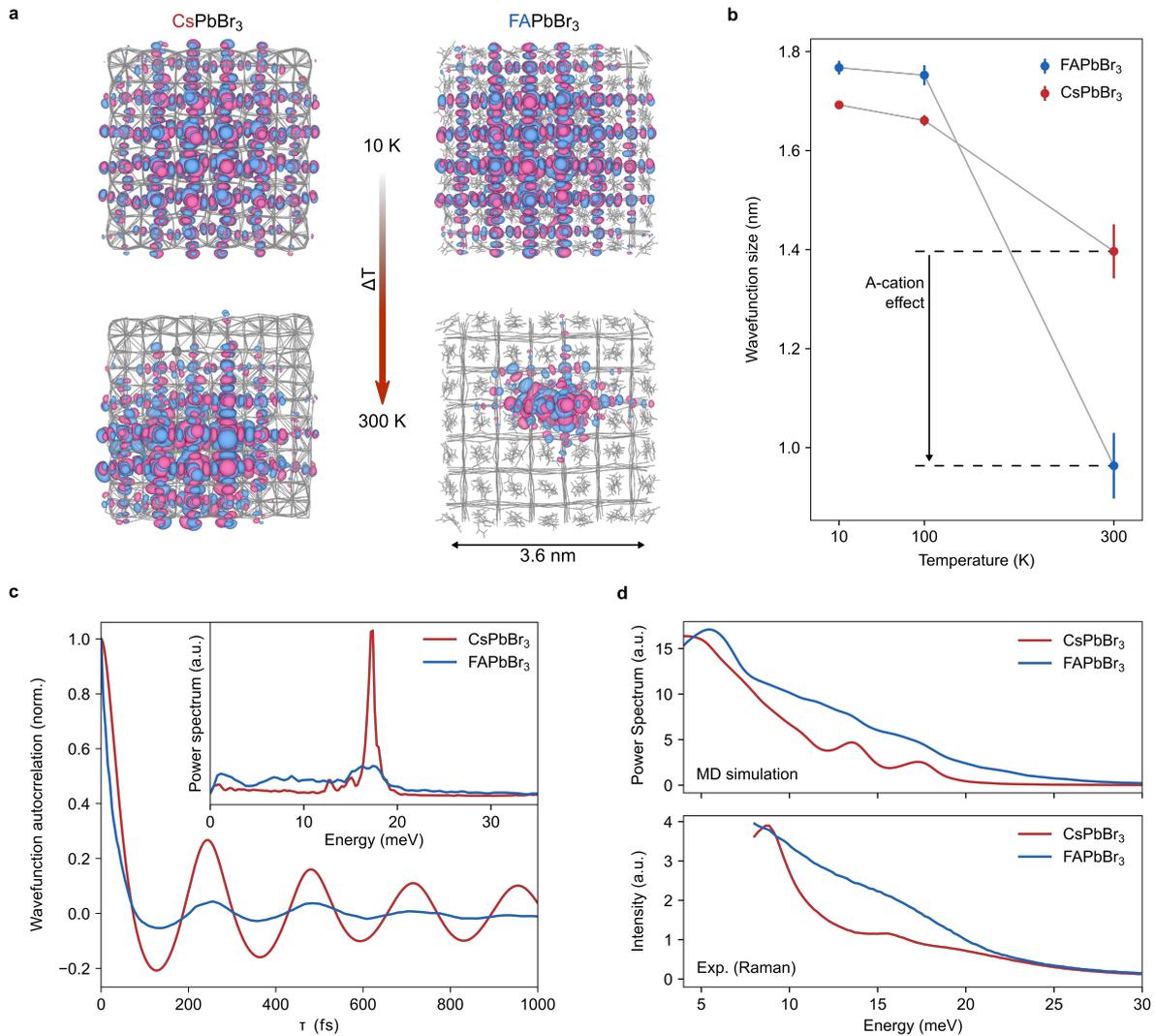

**Figure 3.** Temperature-induced dynamic wavefunction localization and structural origin. (a) Representative snapshots of HOMO wavefunctions obtained from ab-initio molecular dynamics (AIMD) simulations of CsPbBr$_3$ and FAPbBr$_3$ QDs with edge lengths of 3.6 nm at 10 and 300 K. (b) Average HOMO wavefunction size from AIMD trajectories between 10 and 300 K. Error bars indicate 95% confidence intervals. (c) Autocorrelation function of the HOMO

wavefunction obtained by AIMD at 300 K normalized to its value at zero delay. Inset: Spectral density of the wavefunction autocorrelation. (d) Vibrational spectral density from AIMD simulations at 300 K (top) and experimental Raman spectra (bottom) of $CsPbBr_3$ and $FAPbBr_3$ QDs.

We now assess the single-photon purity of similarly-sized (9.9(1.2) nm) $CsPbBr_3$ QDs and $FAPbBr_3$ QDs) in single-particle PL measurements. In CsPbBr3 QDs, $g^2(0)$ values are strongly size-dependent and increase drastically with increasing QD size.[18] Figure 4a shows a representative second-order photon-photon correlation ($g^2(τ)$) of a $CsPbBr_3$ QD. The incomplete antibunching of $g^2(0)$=0.29 translates into a single-photon purity of only 71%, with a high probability of biexciton emission. This observation is consistent with the behaviour of traditional semiconductors for which QD-size confinement is the major knob for tuning multiexciton quenching via Auger-Meitner recombination. When size confinement is progressively lost upon increasing the QD size, so is the Coulomb interaction mediating the Auger-Meitner-recombination of multi-excitons. Consequently, for large QDs with a size exceeding the exciton Bohr diameter, the radiative decay of multi-excitons may increasingly become a competitive channel, hereby compromising the single-photon purity. Several reports have indeed shown that large $CsPbBr_3$ QDs lose their characteristic antibunching due to multi-photon emission from biexcitons.[18,61,62]

In contrast, despite similarly weak QD-size confinement, $FAPbBr_3$ QD achieves a drastically higher single-photon purity of 96% ($g^2(0)$=0.04) (Figure 4b). Figure 4c shows the size-dependence of the single-photon purity including data from >150 single QDs as well as previously published results from our laboratory.[18] $FAPbX_3$ QDs exhibit systematically smaller $g^2(0)$ than comparably size-confined $CsPbBr_3$ QDs supporting the postulated enhancement of quantum confinement in the cubic $FAPbBr_3$ QDs.

In a weak excitation density regime, $g^2(0)$ corresponds to the ratio of the biexciton QY to the exciton QY.[19] Given identical exciton QYs in $CsPbBr_3$ and $FAPbBr_3$ QDs (88% and 89%, respectively), our experiments point to quenched biexciton QYs in $FAPbBr_3$ QDs either (i) due to a speed-up of non-radiative Auger-Meitner recombination of the biexciton and/or (ii) due to slower radiative decay of the biexciton. To disentangle the two contributions, we selectively probe the biexciton emission using heralded single-particle spectroscopy (details in the Supporting Information).[63] Figure 4d displays a representative biexciton PL decay trace of a $CsPbBr_3$ QD with a biexciton lifetime of 0.8 ns. In a representative $FAPbBr_3$ QD, the biexciton PL decays significantly faster with a lifetime of just 0.4 ns (Figure 4e) despite a somewhat slower radiative decay of the exciton (Figure S9). Furthermore, the statistics from 77 QDs show that, while both mechanism (i) and (ii) contribute to quenched biexciton QYs, mechanism (i), i.e., an accelerated non-radiative recombination, is the main lever in quenching biexcitons and enhancing single-photon purity in $FAPbBr_3$ QDs (Figure 4f).

Our AIMD simulations uncovered that disorder-induced wavefunction localization is related to the polymorphism that is most pronounced in pseudo-cubic perovskites. Cubic $APbBr_3$ QDs can also be obtained for the organic cations, such as methylammonium (MA) and aziridinium (AZ). Indeed, weakly size-confined $MAPbBr_3$ QDs and $AZPbBr_3$ QDs exhibit an average single-photon purity of 90(5)% and 89(3)%, comparable to $FAPbBr_3$ QDs of similar size confinement (Figure S10,11).[37,60]

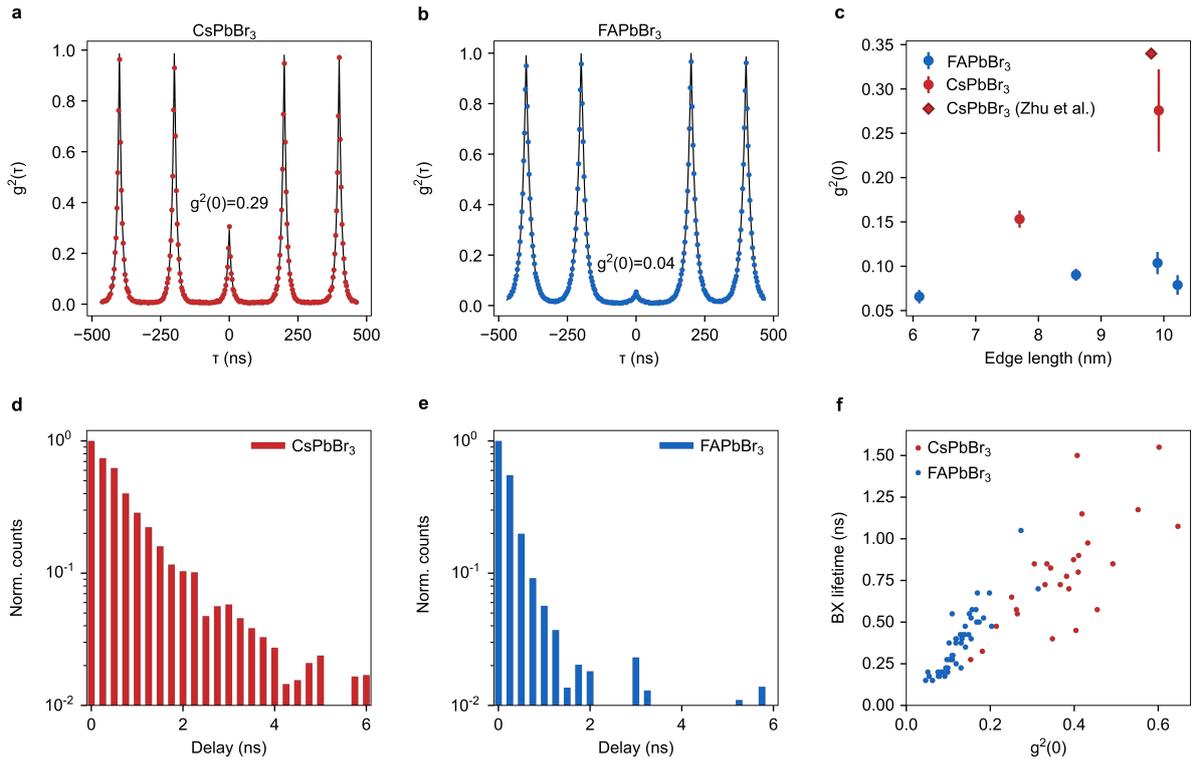

**Figure 4.** Heralded single-particle spectroscopy reveals suppressed multi-photon emission. (a,b) Second-order photon-photon correlation functions $g^2(\tau)$ of similarly size-confined CsPbBr$_3$ (a) and FAPbBr$_3$ (b) QDs with relative sizes of 1.78 and 1.42 times the Bohr diameter $d_B$. A strong "anti-bunching" dip in the coincidences at zero delay time ($g^2(0)$) indicates high single photon purity and a low biexciton quantum yield. (c) Size-dependence of $g^2(0)$ for various APbX$_3$ (A=Cs/FA, X=Br/I) QDs. The relative size is expressed in terms of $d_B$. Error bars indicate 95% confidence intervals. Datapoints for CsPbX$_3$ were extracted from ref [18]. (d,e) Single-particle biexciton PL decay traces obtained for similarly size-confined CsPbBr$_3$ (d) and FAPbBr$_3$ (e) QDs. (f) Single-particle biexciton lifetimes as a function of $g^2(0)$. Enhanced single-photon purity (equivalent to lower $g^2(0)$) correlates with smaller biexciton lifetimes, i.e. quenched multi-exciton emission.

Efficiently quenching multi-excitons via phonon-driven wavefunction confinement opens a new and straightforward avenue toward (i) stable, (ii) bright, and (iii) color-tunable single-photon sources - all enabled by the here-introduced route to circumvent the previous need for QD downsizing to achieve high single-photon purity.

Previously, stability concerns predominantly related to the QD surface, with smaller QDs of high surface-to-volume ratios exhibiting particularly poor photostability. Photodegradation has been attributed to surface-matrix reactions, eventually leading to a shrinkage of the QD core that manifests itself in a dynamical spectral blue-shift and a reduction in PL intensity.[64] Indeed, an individual QD from a large $FAPbBr_3$ QD sample (10.1(9) nm) exhibits extreme photostability with spectrally stable emission during continuous irradiation for 75 minutes, see Figure 5a. Figure 5b,c display the $g^2(\tau)$ measured before and after irradiation of 1 hour, respectively. Strong anti-bunching with $g^2(0)$ of 7% and 5% conveys that single-photon purity does not deteriorate over the course of continuous irradiation for 1 hour. Moreover, the QD displays only very weak blinking, on par with sophisticated core-shell structures[9,10] and remains largely unchanged. Before irradiation, we observe that that the QD spends 98% in its bright (ON) state, while after 1h irradiation an ON fraction of 94% is retained (Figure S12).

Several quantum-engineering applications based on room-temperature single-photon emitters specifically demand bright emitters. For example, quantum key distribution generally requires emitters with high brightness at a single-photon purity >90%.[4] Figure 5d displays the power-dependent PL intensity (top panel) and single-photon purity (bottom panel) of an individual large and weakly confined QD at a laser repetition rate of 10 MHz. We extract a saturation power density $P_{sat}$ of 70 W/cm$^2$ and a maximum brightness of 1.1 Mcps suggesting that the radiative rate limits the brightness (details in Supporting Information). At a low excitation power (0.1 $P_{sat}$), strong anti-bunching attests to a single-photon purity of 95% (Figure 5e). An increase of the excitation to 1.5 $P_{sat}$ only reduces the single-photon purity to 92 % (Figure 5f). In quantum emitters such as defects or molecules, single-photon purity in saturation can be deteriorated by strong background emission at high laser power. However, large perovskite QDs do not suffer from this limitation owing to their orders of magnitude larger absorption cross sections. This is illustrated by the agreement between the measured power dependence of the single-photon purity and the background-free prediction for a biexciton PLQY of 5% of the exciton PLQY (bottom panel; Figure 5d).

Lastly, extending high single-photon purity from small to large FAPbX$_3$ QDs also allows spectral fine-tuning of single-photon emission. Together with the already available compositional tuning through the halide identity, FAPbX$_3$ QDs can now be precision-engineered to deliver high single-photon purity across the entire visible range, for example blue-emitting FAPb(Br/Cl)$_3$ (Figure 5g; 9.2(8) nm edge length), green-emitting FAPbBr$_3$ (Figure 5h, 10.2(1.2) nm edge length), and far-red emitting FAPbI$_3$ (Figure 5i, 11.3(1.3) nm edge length). All QDs display strong antibunching certifying single-photon purities >90%. Blue-emitting single-photon sources receive increasing interest for underwater communication.[65,66] Meanwhile, reaching the far-red and near-infrared spectrum with FAPbI$_3$ QDs is particularly relevant due to reduced losses in fiber-optic communication[2] and biological tissue.[67]

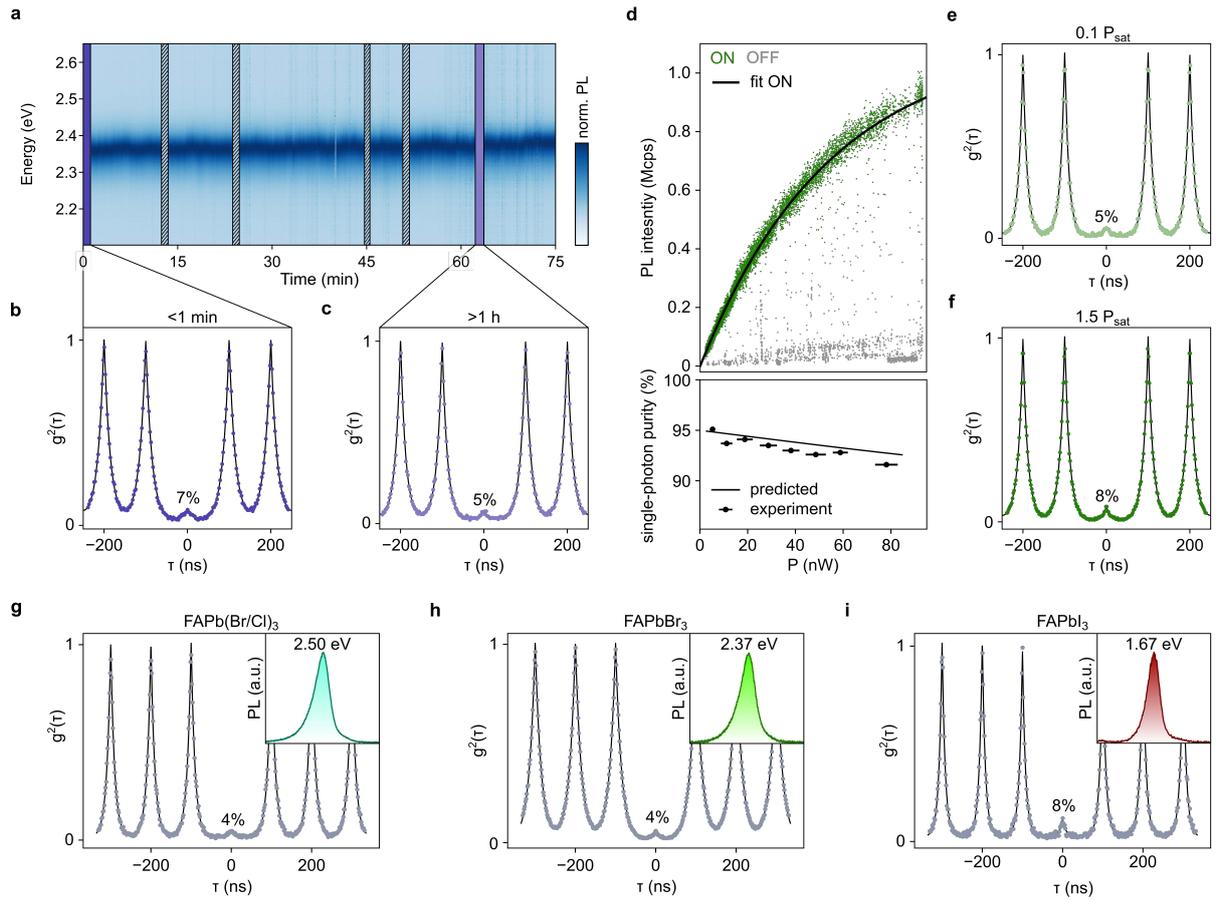

**Figure 5.** FAPbX$_3$ (X=Cl,Br,I) QDs as stable, bright and spectrally tuneable single-photon sources. (a) PL spectra series of a highly photostable FAPbBr$_3$ QD during >1h of continuous irradiation. Shaded areas correspond to coincidence measurements (g$^2$(τ)). (b,c) Measured (green data points) and fitted (black lines) g$^2$(τ) of the QD before (b) and after (c) irradiation for one hour. (d) Top: Excitation power dependence of the PL intensity of an individual FAPbBr$_3$ QD pumped at a repetition rate of 10 MHz. Datapoints corresponding to the QDs bright ON state (blue dots) were fitted by a saturation model (red line) to extract a maximum brightness of 1.1 Mcps and a saturation power density of 70 W/cm$^2$. Bottom: Single-photon purity as a function of excitation power. The grey curve indicates the ideal and background-free behaviour of an ideal QD with a biexciton QY of 0.05 times the exciton QY. Error bars indicate the integrated range of excitation power. (e,f) Measured (blue data points) and fitted (black lines) g$^2$(τ) of the QD recorded at 10% (e) and 150% (f) of the saturation power. (g-i) Measured (blue data points) and fitted (black lines) g$^2$(τ) and PL spectrum (inset) of (g) a FAPb(Br/Cl)$_3$ QD emitting blue single-photons (PL center at 2.50 eV, 495 nm), (h) a FAPbBr$_3$ QD emitting green

single-photons (PL center at 2.36 eV, 525 nm) and (i) a FAPbI$_3$ QD emitting red single-photons (PL center at 1.67 eV, 742 nm).

In summary, we discovered a strategy to enhance quantum confinement in colloidal QDs and improve their single-photon purity without the need for QD downsizing. Caused by anharmonic vibrations that couple to electronic degrees of freedom, dynamic disorder induces wavefunction confinement which increases with temperature and can be tuned through the control of crystal vibrations. In our implementation, wavefunction confinement is realized by deploying organic A-site cations in lead halide perovskite QDs due to their tendency of the resulting pseudo-cubic structures to form polymorphous crystals. Phonon-induced wavefunction localization presents a counterintuitive but beneficial effect of strong exciton-phonon coupling, which is quite often considered detrimental in optoelectronic materials. Owing to the resulting purified single-photon emission in large QDs, we can report competitive brightness, photostability and single-photon purity while retaining the broad spectral tuneability that colloidal QDs stand out for amongst room-temperature single-photon sources. These characteristics are expected to further improve under more favourable conditions such as resonant excitation,[68,69] temporal and spectral filtering,[7,69,70] or integration into optical microcavities and nano antennae.[71,72] Taken together, large organic-inorganic QDs are a versatile and scalable platform of room-temperature single-photon sources.

**Acknowledgements.** This publication was created as part of NCCR Catalysis (grant number 180544), a National Centre of Competence in Research funded by the Swiss National Science Foundation. This work was supported by the Weizmann-ETH Zurich Bridge Program. The project was also partially supported by the European Union's Horizon 2020 program, through a FET Open research and innovation action under Grant Agreement No. 899141 (PoLLoC) and by the Air Force Office of Scientific Research under award number FA8655-21-1-7013.

**Supporting Information.** Additional details on computational methods, quantum dot synthesis, Raman spectroscopy, X-ray diffraction, and single-dot experiments (PDF). Video of the wavefunction density in molecular dynamics simulations of a 3.6 nm $FAPbBr_3$ (MP4) and $CsPbBr_3$ (MP4) QD at 300 K.